\documentclass[aps,prl,twocolumn,showpacs,nofootinbib,superscriptaddress]{revtex4} 

\usepackage{graphicx,longtable,epsfig} %
\usepackage{dcolumn}
\usepackage{mathptmx}
\usepackage{amsmath}
\usepackage{epstopdf}
\usepackage[pdftex]{hyperref}
\usepackage{sidecap}
\sidecaptionvpos{figure}{c}
\usepackage[sort&compress]{natbib}

\newcommand{\beqy}{\begin{eqnarray}}
\newcommand{\eeqy}{\end{eqnarray}}
\newcommand{\bmlet}{\begin{subequations}}
\newcommand{\emlet}{\end{subequations}}
\newcounter{saveeqn}

\def\gsimeq{\,\,\raise0.14em\hbox{$>$}\kern-0.76em\lower0.28em\hbox  
{$\sim$}\,\,}  
\def\lsimeq{\,\,\raise0.14em\hbox{$<$}\kern-0.76em\lower0.28em\hbox  
{$\sim$}\,\,}  

\begin{document}

\title{Evidence for dipole nature of the low-energy $\gamma$ enhancement in $^{56}$Fe}

\author{A.~C.~Larsen}
\email{a.c.larsen@fys.uio.no}
\affiliation{Department of Physics, University of Oslo, N-0316 Oslo, Norway}
\author{N.~Blasi}
\affiliation{INFN, Sezione di Milano, Milano, Italy}
\author{A.~Bracco}
\affiliation{INFN, Sezione di Milano, Milano, Italy}
\affiliation{Dipartimento di Fisica, University of Milano, Milano, Italy}
\author{F.~Camera}
\affiliation{INFN, Sezione di Milano, Milano, Italy}
\affiliation{Dipartimento di Fisica, University of Milano, Milano, Italy}
\author{T.~K.~Eriksen}
\affiliation{Department of Physics, University of Oslo, N-0316 Oslo, Norway}
\author{A.~G\"{o}rgen}
\affiliation{Department of Physics, University of Oslo, N-0316 Oslo, Norway}
\author{M.~Guttormsen}
\affiliation{Department of Physics, University of Oslo, N-0316 Oslo, Norway}
\author{T.~W.~Hagen}
\affiliation{Department of Physics, University of Oslo, N-0316 Oslo, Norway}
\author{S.~Leoni}
\affiliation{INFN, Sezione di Milano, Milano, Italy}
\affiliation{Dipartimento di Fisica, University of Milano, Milano, Italy}
\author{B.~Million}
\affiliation{INFN, Sezione di Milano, Milano, Italy}
\author{H.~T.~Nyhus}
\affiliation{Department of Physics, University of Oslo, N-0316 Oslo, Norway}
\author{T.~Renstr{\o}m}
\affiliation{Department of Physics, University of Oslo, N-0316 Oslo, Norway}
\author{S.~J.~Rose}
\affiliation{Department of Physics, University of Oslo, N-0316 Oslo, Norway}
\author{I.~E.~Ruud}
\affiliation{Department of Physics, University of Oslo, N-0316 Oslo, Norway}
\author{S.~Siem}
\affiliation{Department of Physics, University of Oslo, N-0316 Oslo, Norway}
\author{T.~Tornyi}
\affiliation{Department of Physics, University of Oslo, N-0316 Oslo, Norway}
\affiliation{Institute of Nuclear Research, MTA ATOMKI, H-4026 Debrecen, Hungary}
\author{G.~M.~Tveten}
\affiliation{Department of Physics, University of Oslo, N-0316 Oslo, Norway}
\author{A.~V.~Voinov}
\affiliation{Department of Physics and Astronomy, Ohio University, Athens, Ohio 45701, USA}
\author{M.~Wiedeking}
\affiliation{iThemba LABS, P.O.  Box 722, 7129 Somerset West, South Africa}

\date{\today}

\begin{abstract}
The $\gamma$-ray strength function of $^{56}$Fe has been measured from proton-$\gamma$ 
coincidences for excitation energies up to $\approx 11$ MeV. 
The low-energy enhancement in the $\gamma$-ray strength function, which was first discovered in 
the ($^3$He,$\alpha\gamma$)$^{56}$Fe reaction, is confirmed with the
($p,p^\prime\gamma$)$^{56}$Fe experiment reported here. Angular distributions of the $\gamma$ rays give 
for the first time evidence that the enhancement is dominated by dipole transitions.

\end{abstract}


\maketitle


Atomic nuclei are microscopic systems governed by the laws of quantum mechanics. To understand such systems,
detailed studies of the accessible quantum-energy levels and their decay properties are vital. 
The $\gamma$-ray strength function ($\gamma$SF) is a measure of the average, reduced $\gamma$-decay probability of the nucleus,
and is considered a fruitful concept at high excitation energies where the level spacing is small (the quasi-continuum region).

Structures in the $\gamma$SF provide information on the underlying nuclear dynamics and degrees of freedom, 
such as the $M1$ scissors mode~\cite{heyde2010,schiller2006,krticka2004} 
and the giant electric dipole resonance (GDR)~\cite{harakeh2000}.
The $\gamma$SF is also indispensable for predicting reaction cross 
sections for the astrophysical nucleosynthesis. 
Specifically, when there is no $(n,\gamma)-(\gamma,n)$
equilibrium, the shape of the $\gamma$SF in the vicinity of the neutron threshold 
plays a crucial role for the $(n,\gamma)$ reaction rates relevant for the 
rapid neutron-capture process (r-process)~\cite{go98,ar07}. 

An enhancement in the $\gamma$SF for $\gamma$ energies below $\approx 4$ MeV has 
been discovered in several $fp$-shell and medium-mass nuclei using the 
Oslo method, such as 
$^{56,57}$Fe~\cite{Fe_Alex} and $^{93-98}$Mo~\cite{Mo_RSF}. 
Recently, the low-energy enhancement (hereby denoted \textit{upbend}) was confirmed  
in a ($d,p\gamma$)$^{95}$Mo experiment~\cite{wiedeking95Mo}, using a different detector setup and
a model-independent method to extract the $\gamma$SF. 
The upbend could induce an
increase of up to two orders of magnitude in the ($n,\gamma$) reaction rates 
in very neutron-rich isotopes~\cite{larsen_goriely}. 
Depending on the actual conditions at the astrophysical r-process site, this
could be of great importance for the r-process~\cite{larsen_goriely}. 
 
Despite the potentially crucial role of the upbend for astrophysics applications,
its extent and origin remains largely unknown. 
In particular, the physical mechanism causing the upbend
is not understood, mainly because information on the multipolarity and electromagnetic character is lacking. 
Only for $^{60}$Ni there are data indicating
that the upbend is due to $M1$ transitions~\cite{Ni_Alex}.
However, $^{60}$Ni might be a special case with only positive-parity states
below excitation energies of $\approx 4.5$ MeV.
Up to now, data on Fe isotopes
are inconclusive regarding the radiation type; neither $E1$, $M1$, or $E2$ radiation could
be excluded (see Fig.~3 in Ref.~\cite{Fe_Alex}). 

Recent theoretical works on the upbend suggest that it is of $E1$ nature and due to 
transitions in the single-(quasi)particle continuum~\cite{litvinova2013}, or of $M1$ type and caused by a reorientation
of high-$j$ neutron and proton spins~\cite{schwengner2013}. 
Apart from the single-particle picture, one could also imagine that strong collective 
transitions might cause such an enhancement, for example rotational ($E2$) or vibrational ($E3$) transitions in the quasi-continuum. 
 
In this Letter, we show new data on the $\gamma$SF of $^{56}$Fe. The present data set from the inelastic scattering
reaction $^{56}$Fe($p,p^\prime\gamma$)$^{56}$Fe yielded high statistics and allowed for a detailed analysis 
of the $\gamma$-ray angular distributions. We present here for the first time results on the multipolarity of 
the upbend. 
To our knowledge, this is also the first time where the angular-distribution analyzing tool has been applied
to primary $\gamma$ transitions with a broad distribution of energies at high excitation energies.
 

 \begin{figure}[hbt]
 \begin{center}
 \includegraphics[clip,width=\columnwidth]{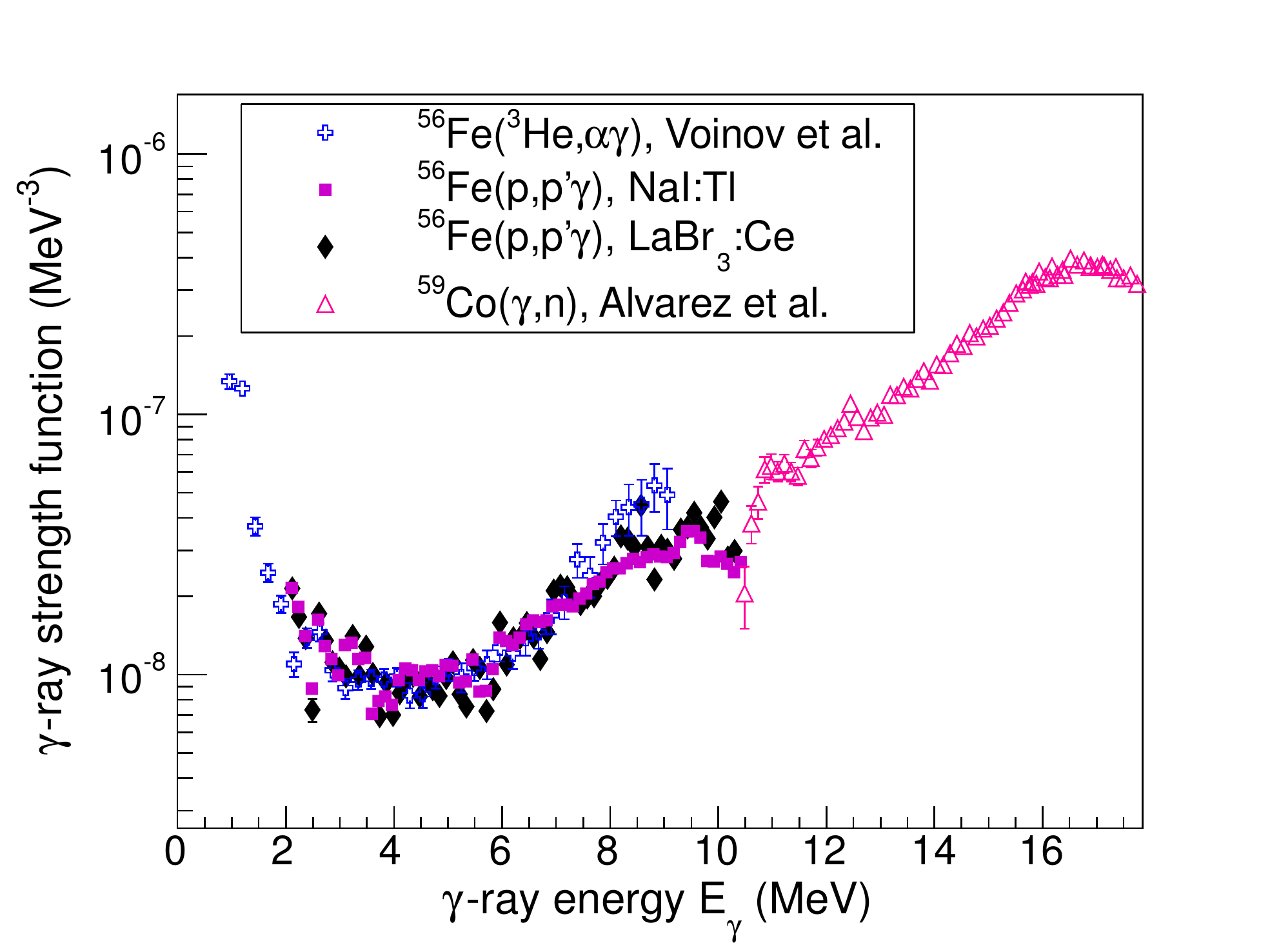}
 \caption {(Color online) Gamma-strength functions of $^{56}$Fe from the present experiment and
    from the ($^3$He,$\alpha\gamma$) data~\cite{Fe_Alex} compared with $^{59}$Co($\gamma,n$) data from
    Ref.~\cite{Alvarez1979}.
 }
 \label{fig:strength}
 \end{center}
 \end{figure}

The experiment was performed at the Oslo Cyclotron Laboratory (OCL), using a 16-MeV proton 
beam with intensity $\approx 0.5$ nA hitting a self-supporting target of 
99.9\% enriched $^{56}$Fe with mass thickness of
2 mg/cm$^2$. Accumulating time was $\approx 85$ hours. The charged ejectiles were measured
with the Silicon Ring (SiRi) particle-detector system~\cite{siri} 
and the $\gamma$ rays with the CACTUS array~\cite{CACTUS}. 
The SiRi system consists of eight $\Delta E - E$ telescopes, where the 
front detector is segmented into eight strips ($\Delta\theta = 2^{\circ}$),
covering scattering angles between 
$40-54^{\circ}$.
In total, SiRi has a solid-angle coverage of $\approx 6$\%. 
Using the $\Delta E-E$ technique, each charged-particle species was identified and a gate was set on the 
outgoing protons. From the reaction kinematics, the proton energy was converted into excitation energy
in the residual nucleus.
 
In this experiment, the CACTUS array contained 22 collimated $5^{\prime\prime} \times 5^{\prime\prime} $ NaI:Tl detectors, and 
six collimated $3.5^{\prime\prime}  \times 8^{\prime\prime} $ LaBr$_3$:Ce detectors~\cite{giaz2013,nicolini2007}.
At the front of the crystals, the conically shaped lead collimators have a radius of 3.5 cm,
and the distance to the target is 22 cm, yielding an internal semi-angle of 9$^\circ$.
The NaI detectors were placed in the CACTUS 
frame with six different angles $\theta$ with respect to the beam axis: 37.4, 63.4, 79.3, 100.7, 116.6, 
and 142.6 degrees, while the LaBr$_3$ crystals covered four angles: 63.4, 79.3, 100.7, and 116.6 degrees.

The $\gamma$ spectra were unfolded using the technique described in Ref.~\cite{gutt6},
but with new response functions from $\gamma$ lines of excited states in $^{13}$C, $^{16,17}$O, 
$^{28}$Si, and $^{56,57}$Fe populated with various inelastic-scattering and transfer reactions. 
Furthermore, the distribution of the primary $\gamma$ rays for each excitation-energy bin (124 keV wide) was determined
from an iterative subtraction technique~\cite{Gut87}. 

 \begin{figure*}[hbt]
 \begin{center}
 \includegraphics[clip,width=1.6\columnwidth]{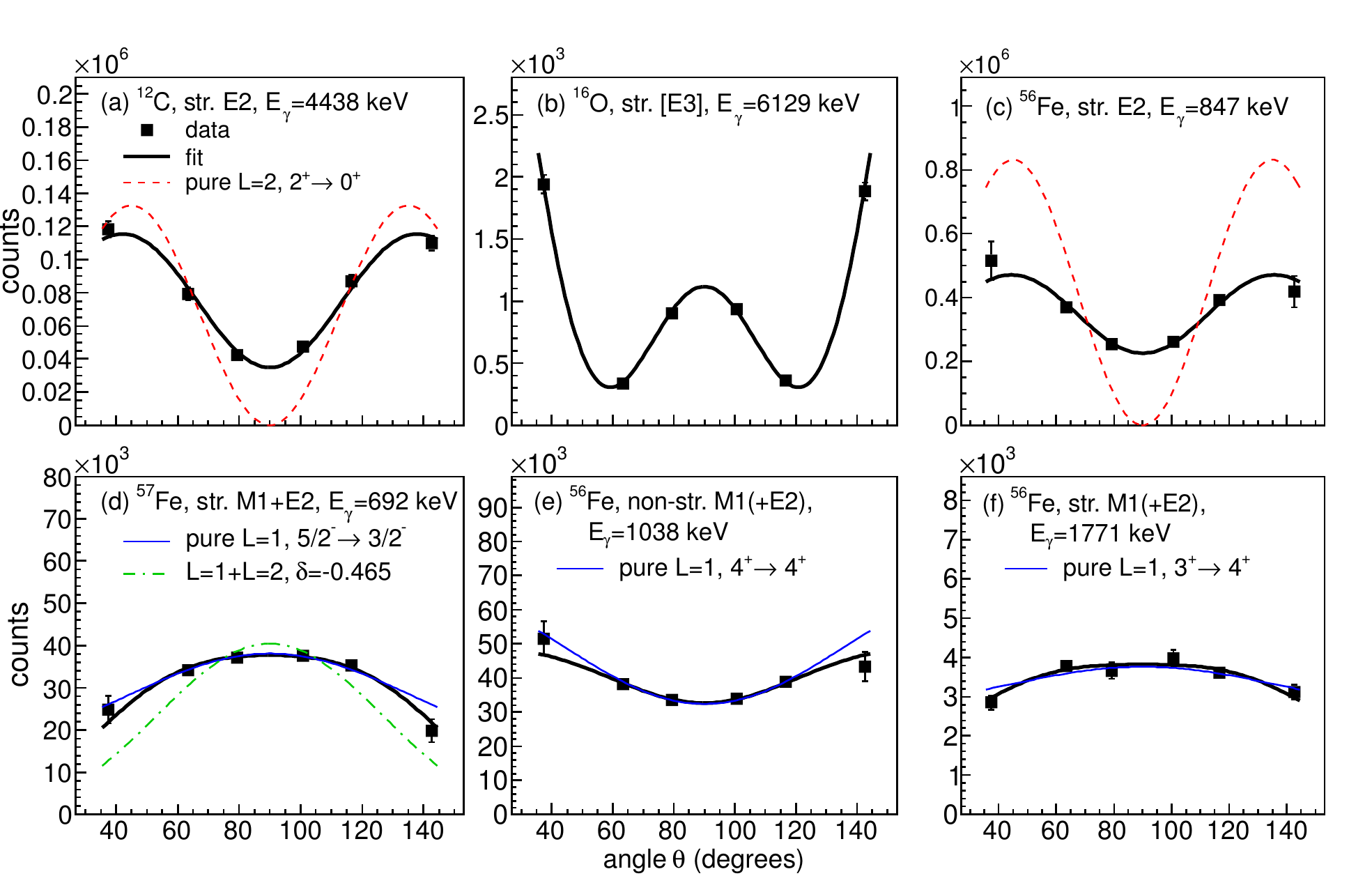}
 \caption {(Color online) Angular distributions of 
    (a) $E2$ in $^{12}$C, 
    (b) [$E3]$ in $^{16}$O, 
    (c) $E2$ in $^{56}$Fe,  
    (d) $M1+E2$ in $^{57}$Fe (mixing ratio $\delta = -0.465(8)$~\cite{ENSDF}, giving an $M1$ fraction of $\approx 82$\%), 
    (e) $M1(+E2)$ in $^{56}$Fe, and  
    (f) $M1(+E2)$ in $^{56}$Fe.  
    All data (black squares) are measured with 
    the NaI detectors. The thick, black lines are
    Legendre fits, the other lines are theoretical 
    distributions~\cite{mateosian1974} with no attenuation (see text). 
 }
 \label{fig:angdist_COFe}
 \end{center}
 \end{figure*}
\begin{table*}[!htb]
\caption{Angular-distribution coefficients of transitions measured in the present experiment (see text).
    The theoretical $a_k^{\mathrm{max}}$ coefficients for complete alignment are taken from Ref.~\cite{mateosian1974}.} 
\begin{tabular}{lcccccccc}
\hline
\hline
$^A X$     & $E$   & $E_\gamma$  & $I_i \rightarrow I_f$     & $XL$    & $a_2^{\mathrm{max}}$ & $a_2$       & $a_4^{\mathrm{max}}$ & $a_4$       \\
		   & (keV) & (keV)       &                           &         &                      &             &                      &             \\
\hline
$^{12}$C   &  4439 & 4438        & $2^+ \rightarrow 0^+$     & $E2$    &  0.714               & 0.55(9)     & $-1.71$              & $-0.77(13)$ \\
$^{16}$O   &  6130 & 6129        & $3^- \rightarrow 0^+$     & $[E3]$  &  $-$                 & 1.85(8)     & $-$                  & $1.91(9)$   \\
$^{56}$Fe  &   847 &  847        & $2^+ \rightarrow 0^+$     & $E2$    &  0.714               & 0.29(18)    & $-1.71$              & $-0.60(13)$ \\
$^{56}$Fe  &  3123 & 1038        & $4^+ \rightarrow 4^+$     &$M1(+E2)$& $0.500$              & 0.31(13)    & 0.00                 &$-0.09(8)$  \\
$^{56}$Fe  &  3856 & 1771        & $3^+ \rightarrow 4^+$     &$M1(+E2)$& $-0.167$             & $-0.33(8)$  & 0.00                 & $-0.11(14)$ \\
$^{56}$Fe  &  4510 & 3663        & $3^- \rightarrow 2^+$     & $(E1)$  & $-0.400$             & $-0.31(16)$ & 0.00                 & $0.07(13)$ \\
$^{56}$Fe  &  5122 & 3037        & $5^- \rightarrow 4^+$     & $(E1)$  & $-0.333$             & $-0.42(15)$ & 0.00                 & 0.20(17)    \\
$^{57}$Fe  &   706 &  692        & $5/2^- \rightarrow 3/2^-$ & $M1+E2$ & $-1.068$             & $-0.69(12)$ & 0.12                 & $-0.18(9)$  \\
\hline
\hline
\end{tabular}
\\
\label{tab:coeff}
\end{table*}

 \begin{figure}[hbt]
 \begin{center}
 \includegraphics[clip,width=1\columnwidth]{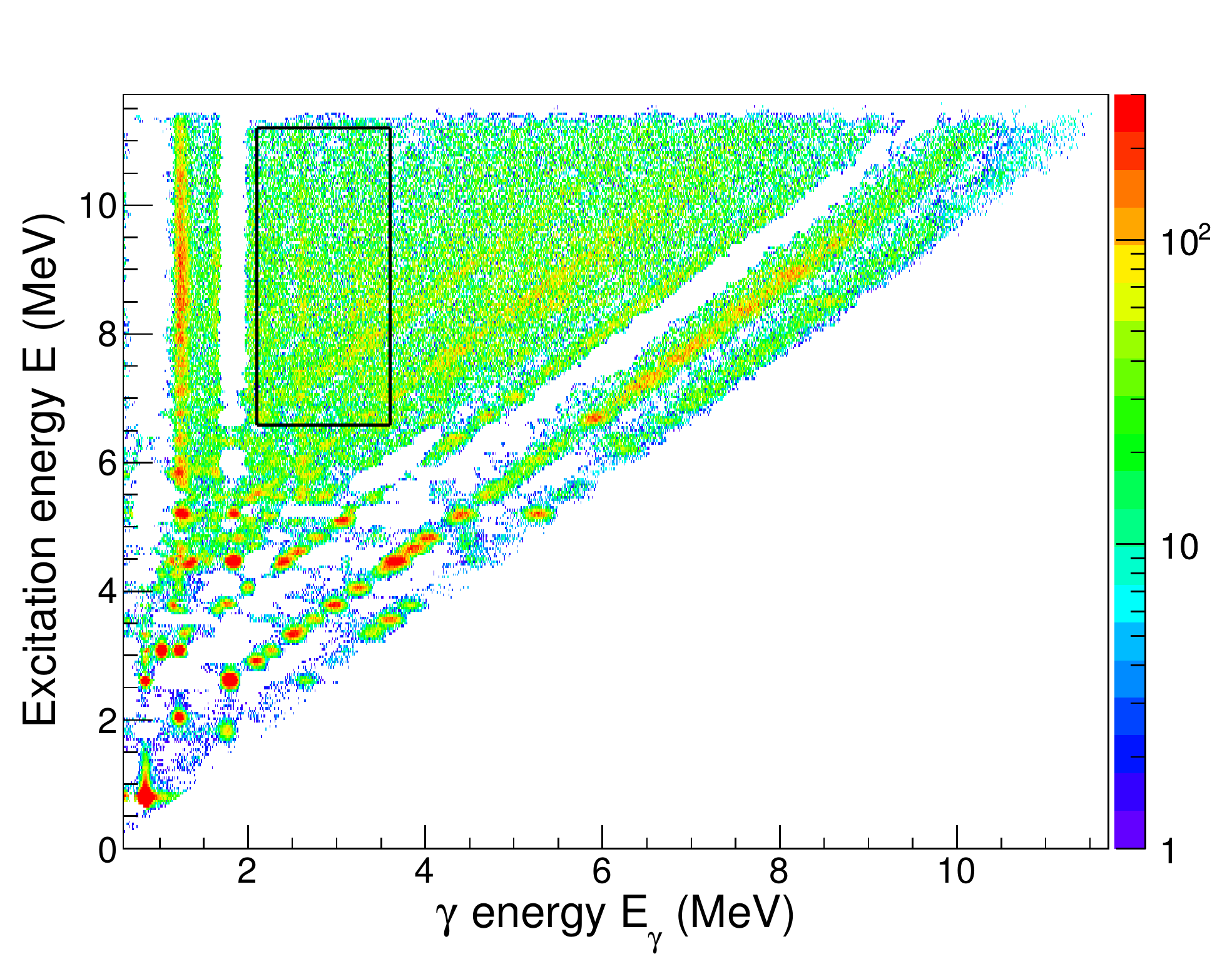}
 \caption {(Color online) Distribution of primary $\gamma$ rays in $^{56}$Fe from the NaI detectors at 79.3$^\circ$.
    The box indicates the region used for the angular distributions of $E_\gamma<3.6$ MeV. 
 }
 \label{fig:fg_Fe}
 \end{center}
 \end{figure}

From the matrix of primary $\gamma$ spectra, we have extracted simultaneously the level density and $\gamma$-transmission 
coefficient for $^{56}$Fe using the least $\chi^2$ method given in Ref.~\cite{Schiller00}.
The absolute value and slope of the level density were determined from discrete levels~\cite{ENSDF} below an excitation energy 
of $E=4$ MeV
and from the comparison to particle-evaporation data~\cite{Voinov2006,Vonach1966}. To get the absolute
value of the $\gamma$-transmission coefficient, we used estimated values from systematics (as there are no experimental values) 
for the neutron-resonance level spacing 
$D_0 = 2500(1250)$ eV and the total, average $\gamma$ width $\left< \Gamma_\gamma \right> = 1500 (750)$ meV, and spin cutoff
parameters from Ref.~\cite{egidy2009}. Assuming that dipole radiation dominates the $\gamma$ decay in the 
quasi-continuum region, the $\gamma$SF is deduced from the $\gamma$-transmission coefficient by
\begin{equation}
f(E_{\gamma}) = {\mathcal T}(E_{\gamma})/2\pi E_{\gamma}^{3},
\label{eq:rsf}
\end{equation}
where $f(E_{\gamma})$ is the $\gamma$SF for $\gamma$ energy $E_{\gamma}$, and ${\mathcal T}(E_{\gamma})$ is
the $\gamma$-transmission coefficient.
The resulting $\gamma$SFs obtained from the LaBr$_3$ and NaI $\gamma$ spectra
are shown in Fig.~\ref{fig:strength}. 

We observe that our new data are
in overall very good agreement with the ($^3$He,$\alpha\gamma$) data of Ref.~\cite{Fe_Alex}. The upbend
is confirmed, using new, higher-resolution detectors and response functions. 
Also, the different reaction type is expected to populate lower initial spins than the 
($^3$He,$\alpha\gamma$) reaction, which has a high cross section for high-$\ell$~pickup~\cite{casten1972}. 
Compared to the ($^3$He,$\alpha\gamma$) experiment, 
the particle-detector resolution has been improved from 400 keV to 90 keV (full-width half maximum), and the $\gamma$-energy
resolution has been improved by more than a factor of 2 for all $\gamma$ energies using the LaBr$_3$ crystals. 
Thus, the upbend is 
clearly independent from 
systematic errors in the detector response and reaction-induced effects.
The difference in strength at high $\gamma$ energies might be due to small variations in the normalization of 
the level density and the new and more precise response functions. 
Also, we see a good match with photo-neutron data on $^{59}$Co~\cite{Alvarez1979}, supporting the chosen values for 
$D_0$ and $\left< \Gamma_\gamma \right>$.

\begin{figure*}[hbt]
 \begin{center}
 \includegraphics[clip,width=1.3\columnwidth]{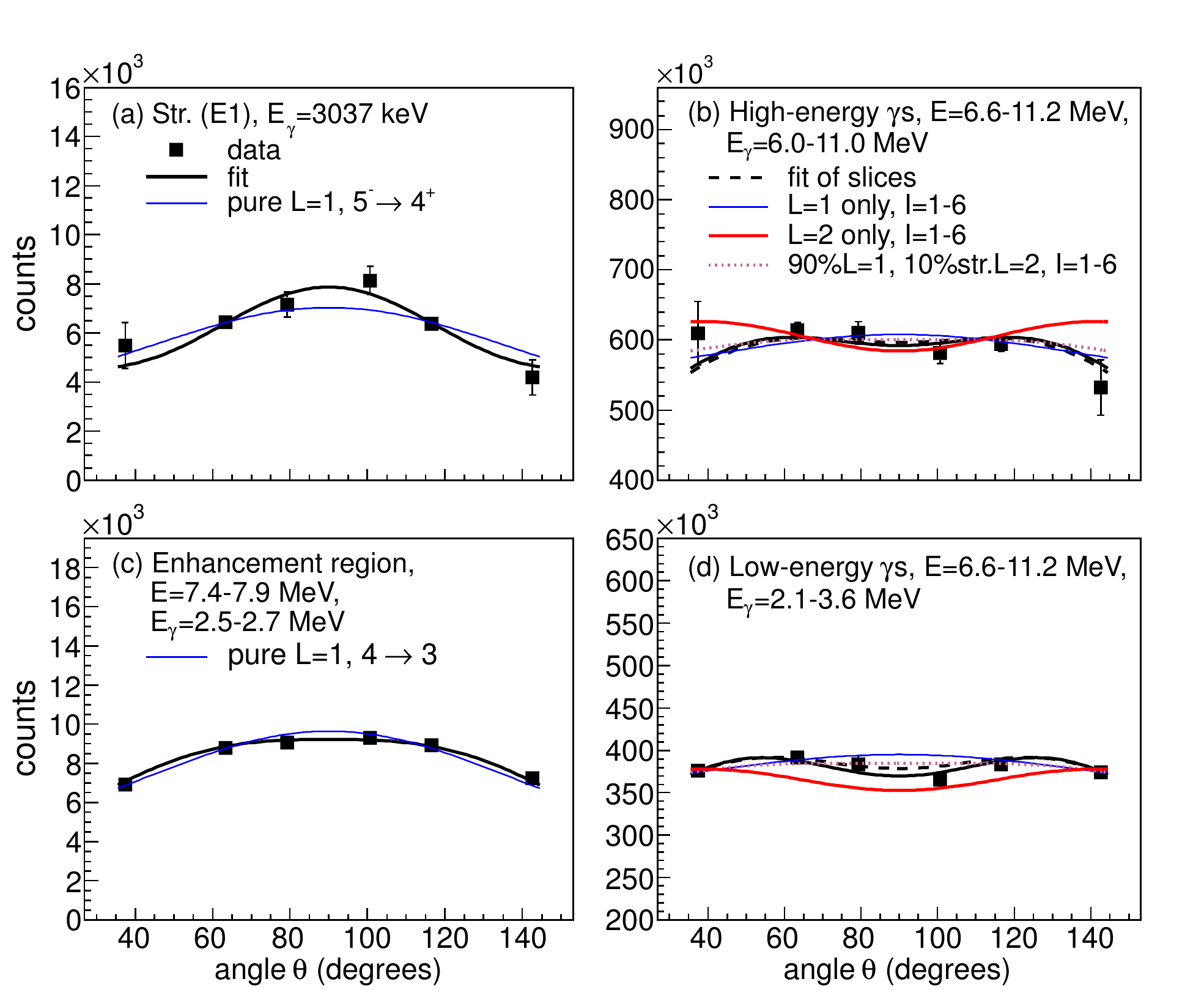}
 \caption {(Color online) Angular distributions from the primary-$\gamma$ matrix of $^{56}$Fe:
 (a) $(E1)$ transition; 
 (b) high-energy primary $\gamma$ rays;
 (c) narrow $E,E_\gamma$ gate in the low-energy $\gamma$ region; 
 (d) low-energy primary $\gamma$ rays.
 The thick, black lines show the fit to the data, and the thin blue line gives theoretical dipole distributions. 
 For (b) and (d) the dashed lines show the angular distributions from the fit of many $E$ slices, 
 the blue (red) lines give the theoretical curves for dipole (quadrupole) transitions from initial spins $I_i=1-6$,
 and the pink, dotted lines are theoretical curves for a mix of $L=1$ and $L=2$ transitions (see text). Note the different
 scale on the y axis for (b) and (d).}
 \label{fig:angdist_upbend}
 \end{center}
\end{figure*}

Making use of the various angles for which the NaI detectors were placed, angular distributions were extracted
by sorting the data into $(E,E_\gamma)$ matrices according to the
angle $\theta$ of the NaI detectors relative to the beam direction. From the intensities as a function of angle, 
we have fitted angular-distribution functions
of the form~\cite{mateosian1974} 
\begin{equation}
W(\theta) = A_0 + A_2 P_2(\cos \theta) + A_4 P_4 (\cos \theta),
\label{eq:legendre}
\end{equation}
where $P_k(\cos \theta)$ is a Legendre polynomial of degree $k$. The LaBr$_3$ detectors were placed at only four angles
and were not used for this analysis, although we note that the shape of the angular distributions for the 
LaBr$_3$ and NaI detectors are in very good agreement for the four overlapping angles.

The normalized angular-distribution coefficients are given by $a_k = Q_k \alpha_k A_k/A_0$, where $Q_k\approx 1$ is the 
geometrical attenuation coefficient due to the finite size of the $\gamma$ detectors, and $\alpha_k$ is the attenuation 
due to partial alignment of the nuclei relative to the beam direction. Errors in the intensities are given by
$\sigma_{\mathrm{tot}} = \sigma_{\mathrm{stat}} + \sigma_{\mathrm{syst}}$, where the statistical errors are estimated with 
$\sqrt{N}$ where $N$ is the number of counts, and the systematic errors are deduced from the relative change in $N$
for each symmetric pair of angles (37.4$^\circ$,142.6$^\circ$), (63.4$^\circ$,116.6$^\circ$), and (79.3$^\circ$,100.7$^\circ$). 
Note that for this high-statistics experiment, the statistical error bars are in general small. However, the systematic 
uncertainties due to partly asymmetric $\gamma$ intensities for the pairs of angles can in some cases be rather large,
which in turn influence the uncertainties in the $a_k$ coefficients.

The resulting angular distributions for the 4.4-MeV $E2$ transition in $^{12}$C and the 
6.1-MeV [E3] transition in $^{16}$O are shown in Fig.~\ref{fig:angdist_COFe} (a) and (b). 
Correspondingly, transitions
in $^{56}$Fe and $^{57}$Fe  are shown in Fig.~\ref{fig:angdist_COFe} (c)$-$(f). 
The extracted angular-distribution coefficients are given in Tab.~\ref{tab:coeff}. 
The stretched dipole, quadrupole and octupole transitions 
are easily distinguished from each other. We also observe that the attenuation due to 
partial alignment is becoming less and less pronounced as the excitation energy increases; in fact, the $a_2^{\mathrm{max}}$
coefficients are in good agreement with the data above $E \approx 3$ MeV in $^{56}$Fe 
(see Fig.~\ref{fig:angdist_COFe} (e) and (f)).

We now turn to the distribution of primary $\gamma$ rays as function of excitation energy.
The matrix of primary-$\gamma$ spectra for $^{56}$Fe is displayed in Fig.~\ref{fig:fg_Fe}.
For $\gamma$ decay in the quasi-continuum below the neutron threshold, 
the $\gamma$SF is dominated by the tail of the GDR. 
In addition, the Giant Magnetic Dipole Resonance (GMDR), has its maximum at typically
$E_\gamma = 8$ MeV~\cite{RIPL}. Thus, the region
of high excitation energy (above $\approx 5-6$ MeV) is expected to be dominated by dipole transitions. 

In the present experiment, the reaction populates a range of initial spins in the quasi-continuum. From the primary transitions
we can clearly identify initial spins up to $6$. The angular distributions represent a 
mix of stretched and non-stretched dipole transitions; if an initial level with spin 4 is populated, it 
might de-excite with a dipole transition to a final level with spin 3, 4, or 5. 
Two of these transitions are stretched and one
is non-stretched, therefore, one expects that on average 2/3 of the transitions are stretched and 1/3 are non-stretched.

The angular distributions for a non-stretched and a stretched $M1(+E2)$ transition in $^{56}$Fe
are shown in Fig.~\ref{fig:angdist_COFe} (e) and (f), while in Fig.~\ref{fig:angdist_upbend} (a)
a stretched ($E1$) transition is displayed. 
The angular distribution of high-energy $\gamma$ rays for $E>6.6$ MeV is shown in Fig.~\ref{fig:angdist_upbend} (b),
and for a narrow gate in the region of the
upbend in Fig.~\ref{fig:angdist_upbend} (c), with a shape
consistent with a stretched dipole (the exact initial and final spin is unknown). 
A theoretical distribution
assuming a $4\rightarrow 3$ transition is shown, using values of 
$a_2^{\mathrm{max}}=-0.357$, $a_4^{\mathrm{max}}=0.0$~\cite{mateosian1974}, to be compared with the values from the fit, 
$a_2=-0.35(4)$, $a_4=-0.10(6)$.
The angular distribution for the whole low-energy region 
(the box in Fig.~\ref{fig:fg_Fe}) is displayed in Fig.~\ref{fig:angdist_upbend} (d), clearly resembling the high-energy part.  

To determine the angular-distribution coefficients for the high-energy $\gamma$ rays and in the region of the upbend, 
we have performed independent fits of Eq.~(\ref{eq:legendre}) to 720-keV wide excitation-energy slices of the primary $\gamma$ matrix. 
Then, a linear fit was performed for all the extracted angular-distribution coefficients, giving $a_2 = -0.07(1)$, $a_4 = -0.09(1)$
and $a_2 = -0.12(3)$, $a_4 = -0.08(3)$ for the low and high-energy $\gamma$ rays, respectively (dashed lines 
in Fig.~\ref{fig:angdist_upbend}). 
The $a_k$ coefficients for the two energy regions are compatible within $1\sigma$, which indicate that the nature of these
$\gamma$ rays is very similar. 
By applying a weight of 2/3 for the stretched and 1/3 for the non-stretched 
known dipole transitions in $^{56}$Fe as given in Tab.~\ref{tab:coeff} ($E_\gamma=1038$, 1771, 3037 and 3663 keV), 
the expected $a_k$ coefficients for the quasi-continuum decay 
are $a_2 = -0.13(7)$ and $a_4 = -0.01(5)$, further supporting that both the low and high-energy $\gamma$ regions 
are dominated by dipole transitions. Based on these findings, we can exclude that the upbend is due to stretched quadrupole ($E2$) or octupole ($E3$) transitions.

We have also considered expected distributions with $a_k^{\mathrm{max}}$ coefficients~\cite{mateosian1974}
 for an initial spin range $I_i=1-6$ and final spins
$I_f=0-7$ for stretched and non-stretched dipole transitions, yielding the distribution shown as a blue line in   
Fig.~\ref{fig:angdist_upbend} (b) and (d). If we assume that there are only quadrupole transtions (stretched and non-stretched,
$I_f=0-8$), 
the fit is much worse and the data are clearly not reproduced (red lines in Fig.~\ref{fig:angdist_upbend} (b) and (d)).
The best reproduction of the experimental angular distributions was found with a 90\% and 10\% weight on the 
dipole and stretched-quadrupole contribution 
($6\rightarrow4, 5\rightarrow3, 4\rightarrow2, 3\rightarrow1$). 
For an increased weight on the quadrupole contribution, or taking non-stretched and $4\rightarrow 6, ...$ quadrupoles into account,
the fit was significantly worse. Therefore, we conclude that $E2$ transitions are of minor importance and that dipole transitions
dominate both the region of the upbend and for the high-energy $\gamma$s. Our findings support the $L=1$ 
assumption applied in Eq.~(\ref{eq:rsf}).

To summarize, we have presented in this Letter a new measurement on the $\gamma$-strength function of $^{56}$Fe. The upbend 
in the strength, which may have profound consequences for r-process reaction rates, is confirmed with an
improved detector setup and response functions, and with a different reaction and beam energy. We have demonstrated that the angular
distribution of the low-energy primary $\gamma$ rays is consistent with a mixture of stretched and non-stretched dipole transitions,
and that quadrupole and octupole transitions are of minor importance. Thus, for the first time, the multipolarity of the upbend 
has been measured and shown to exhibit predominantly a dipole character.

\begin{acknowledgments}
A.~C.~L. gratefully acknowledges funding of this research from the Research Council of Norway, project grant no. 205528. 
M.~W. acknowledges support from the National Research Foundation of South Africa.
We would like to give special thanks to 
E.~A.~Olsen, J. C. M\"{u}ller, A.~Semchenkov, and J.~C.~Wikne for providing the high-quality beam and excellent experimental conditions. 
\end{acknowledgments}

\end{document}